\documentclass[prb,twocolumn,aps,showpacs]{revtex4}
\usepackage{graphicx,bm}
\newcommand{\mscr}[1]{{\mbox{\scriptsize #1}}}
\begin{document}
\title{Hyperfine-mediated transitions between a Zeeman split doublet
in GaAs quantum dots: The role of the internal field}
\author{Sigurdur I.\ Erlingsson, Yuli V.\ Nazarov}
\affiliation{Delft University of Technology, Department of Applied Physics,
         Lorentzweg 1, 2628 Delft, The Netherlands}
\date{\today}
\begin{abstract}
We consider the hyperfine-mediated transition rate between
Zeeman split spin states of the lowest orbital level in a GaAs
quantum dot.  
We separate the hyperfine Hamiltonian into a part which
is diagonal in the orbital states and another one which mixes different
orbitals.  The diagonal part gives rise to an 
effective (internal) magnetic field which, in addition to an external 
magnetic field, determines the Zeeman splitting.  
Spin-flip transitions in the dots are induced by the
orbital mixing part accompanied by an emission of a phonon.  We
evaluate the rate for different regimes of applied magnetic field
and temperature.  The rates we find are bigger that the spin-orbit
related rates provided the external magnetic field is sufficiently
low. 
\end{abstract}
\pacs{71.23.La, 71.70.Jp}
\maketitle
\section{Introduction}
The electron spin states in bulk semiconductor and heterostructures have
attracted much attention in recent years. Experiments indicate very long
spin decoherence times and small transition rates between states of
different spin \cite{kikkawa98:4313,ohno00:817,fujisawa01:R81304}. These
promising results have motivated proposals for information processing based
on electron spins in quantum dots, which might lead to a realization of a
quantum computer \cite{loss98:120,burkard99:2070}.

A quantum dot is region where electrons are confined. The energy spectrum is
discrete, due to the small size, and can display atomic-like properties \cite
{ashoori96:413,tarucha96:3613}. Here we will consider quantum 
dots in GaAs-AlGaAs heterostructures. The main reasons for studying them
are that relevant quantum dots are fabricated in such structures and
Ga and As nuclei have a substantial hyperfine interactions with 
the conduction electrons.
There are two main types of gate controlled dots in these systems, so-called
vertical and lateral dots\cite{sohn97:105}. They
differ in transverse confinement, which is approximately a triangular well
and a square well for the lateral and vertical dots, respectively.

Manipulation the electron spin while maintaining phase coherence
requires that it should be 
relatively well isolated from the environment. Coupling a
quantum dot, or any closed quantum system, to its environment can cause
decoherence and dissipation. One of the measures of the strength of
the coupling to 
the environment are the transition rates, or inverse lifetimes, between the
quantum dot states. 
In GaAs there are two main mechanisms that can cause finite lifetimes of
spin states.  These are the spin-orbit interaction and the hyperfine
interaction with the surrounding nuclei.  
If a magnetic field is applied  the change in Zeeman energy
accompanying a spin-flip has to be absorbed by phonon emission.  
For two electron quantum dots, where the transitions are between
triplet and singlet spin states, both
spin-orbit\cite{khaetskii00:12639,khaetskii00:470} and
hyperfine\cite{erlingsson01:195306} mediated transitions have been
studied.  In both cases the transition rates are much smaller than the
usual phonon rates, the spin-orbit rate being the higher except 
when the excited singlet and triplet states
cross.\cite{erlingsson01:195306}  For low magnetic  fields, i.e.\ 
away from the  single-triplet transition, the energy of the emitted
phonon can be quite large and the transitions involve deformation
phonons rather than piezoelectric ones.  

If there is odd number of electrons in the dot the ground-state is
usually a spin doublet so that energy change associated with the
spin-flip is the electron Zeeman energy.
Owing to the small $g$-factor in GaAs this energy is rather 
small compared to the level spacing and the dominating phonon mechanism
is due to piezoelectric 
phonons.  Recently spin-orbit mediated spin-flip transitions between
Zeeman levels was investigated\cite{khaetskii01:27}.  Due to Kramer's
degeneracy the transition amplitude for the spin-flip is proportional
to the Zeeman splitting.  This results in a spin-orbit spin-flip rate
proportional to the fifth power of the Zeeman splitting.

In this publication we consider hyperfine mediated transitions
between Zeeman split levels in a quantum dot.  
The transition amplitude remains finite at zero external magnetic
field, giving a spin-flip rate (Eq.\ (\ref{eq:finalRate})
) that is proportional to the cube of the Zeeman splitting. 
The cause of this is an internal magnetic field
due to the hyperfine interaction.
We consider the important concept of internal magnetic field in some
detail. 
Since the parameters of hyperfine interaction
between conduction band electrons and underlying nuclei in GaAs have been
extensively investigated \cite{paget77:5780,meier84:381}, including the
Overhauser effect and spin-relaxation in GaAs/AlGaAs heterostructures \cite
{dobers88:1650,berg90:2563,vagner82:635,barret95:5112,tycko95:1460} , we are
able now to predict the typical time scale for this process in particular
quantum dot geometries. 

Upon completion of this work we learned about recent results of
Khaetskii, Loss and Glazman \cite{khaetskii02:1}. They consider essentially the
same situation and model and obtain {\em electron spin decoherence}
without considering any mechanism of dissipation.  This is in clear
distinction from the present result for {\em spin-flip rate} that
requires a mechanism of dissipation, i.e.\ phonons.

The rest of the paper is organized as follows:
in section \ref{sectionII} the model used is introduced  in addition
to the  basic assumptions and approximations used, section
\ref{sectionIII} deals with the internal magnetic field due to the
hyperfine interaction and section \ref{sectionIV} contains the
derivation of the transition rates.  Finally, in section \ref{sectionV}
the results are discussed.
\section{Model and assumptions}
\label{sectionII}
We consider a quantum dot embedded in a AlGaAs/GaAs
heterostructure.  The details of the quantum dot eigenstates are not
important for now, it suffices to say that the energy spectrum
is discrete and the wave functions are localized in space.
The spatial extension of the wavefunction in the lateral and
transverse direction (growth direction) are denoted with  $\ell$ and
$z_0$, respectively.  
Quantum dots in these heterostructures are formed at a GaAs/AlGaAs
interface, where the confining potential is very strong so that 
$\ell \gg z_0$.  
We define the volume occupied by an electron as $V_\mscr{QD}=\pi
\ell^2 z_0$. 
The Hamiltonian of the quantum dot can be written on the form
\begin{equation}
H_0=\sum_l 
\biglb (
\varepsilon_l +g\mu_B \bm{B}\cdot \hat{\bm{S}}
\bigrb )
| l\rangle \langle l| ,
\label{eq:electronHamiltonian}
\end{equation}
where $\varepsilon_l$ are the eigenenergies
which depend on the
structure of the confining potential and the applied magnetic field.
The magnetic field also couples to the electron spin via the Zeeman
term.  

Since the $\Gamma$ point of the conduction band in GaAs is mainly
composed of $s$ orbitals the dipole interaction with the nuclei vanishes
and the hyperfine interaction can be described by the usual contact term
\begin{equation}
H_\mscr{HF}=A\hat{\bm{S}}\cdot \sum_k \hat{\bm{I}}_k \delta(\bm{r}-\bm{R}_k),
\label{eq:hyperfineHamiltonian}
\end{equation}
where $\hat{\bm{S}}$ ($\hat{\bm{I}}_k$) and $\bm{r}$ ($\bm{R}_k$)
denotes, respectively, 
the spin and position of the electron ($k$th nuclei).  The delta
function indicates the point-like nature of the contact interaction
will result in a position dependant coupling.  The coupling constant $A$
has the dimension Volume$\times$Energy.  
To get a notion of the related energy scale, it is straightforward
to relate $A$ to the energy 
splitting of the doublet for a fully polarized nuclear system,
\begin{equation}
E_n=A C_n I,
\end{equation}
$C_n$ being density of nuclei and $I$ is the spin of
a nucleus. 
In GaAs this energy is $E_n\approx 0.135$\,meV, which
corresponds to a magnetic field of about 5\,T.\cite{dobers88:1650} 
For a given quantum dot geometry the number of nuclei
occupying the dot is defined as $N_\mscr{QD}=C_n V_\mscr{QD}$.
Since the Ga and As nuclei have the same spin and their 
coupling constants are comparable, we will assume that all the nuclear
sites are characterized by the same hyperfine coupling $A$, and the
$C_n \approx a_0^{-3}$ where $a_0$ is the lattice spacing.
For realistic quantum dots $N_\mscr{QD} \approx 10^4-10^6$, and it is
therefore an important big parameter in the problem.

The coupling between the electron and the phonon bath is represented
with
\begin{equation}
H_\mscr{ph}=\sum_{\bm{q},\nu} \alpha_\nu(\bm{q})(b_{\bm{q},\nu}e^{i
\bm{q}\cdot\bm{r} }+b^\dagger_{\bm{q},\nu}e^{-i
\bm{q}\cdot\bm{r}}),
\label{eq:electronPhononHamiltonian}
\end{equation}
where $b^\dagger_{\bm{q}\nu}$ and $b_{\bm{q}\nu}$ are creation and
annihilation operator for the phonon mode with wave-vector $\bm{q}$ on
brach $\nu$.  
In GaAs there are two different coupling
mechanisms, deformation ones and 
piezoelectric ones.  For transitions between Zeeman split levels in GaAs,
i.e.\ low energy emission, the most effective phonon mechanism is due
to piezoelectric phonons.  We will assume that the heterostructure
is grown in the [100] direction. This is the case for almost all dots and
it imposes important symmetry relations on the coupling coefficient.
The square of the coupling coefficient for
the piezoelectric phonons is then given by (see Ref.\ \onlinecite{price81:217})
\begin{equation}
\alpha^2_\nu(\bm{q})=\frac{(eh_{14})^2\hbar}{2\rho c_\nu V
q}A_\nu(\theta)
\end{equation}
where $(eh_{14})$ is the piezoelectric coefficient, $\rho$ is the mass
density, $c_\nu$ is the speed of sound of branch $\nu$, $V$ is
normalization volume and we have defined $\bm{q}=q(\cos \phi \sin
\theta,\sin \phi \sin \theta, \cos \theta )$ and the $A_\nu$'s are the
so-called anisotropy functions, see appendix
\ref{app:integral}.

\section{Internal magnetic field}
\label{sectionIII}
In this section we will introduce the concept of the effective magnetic
field, the {\em internal field}, acting on the electron due to the hyperfine interaction.  
This internal field is a semiclassical approximation to the nuclear
system, this approximation being valid in the limit of large number of
nuclei, $N_\mscr{QD} \gg 1$.
If the nuclei are noticeably polarized, this field almost coincides
with the Overhauser field that represents the average nuclear
polarization.   It is important that the internal field persists even
at zero polarization giving rise to doublet splitting of the order
$E_n N_\mscr{QD}^{-1/2}$.

First we write the Hamiltonian in Eq.\
(\ref{eq:hyperfineHamiltonian}) 
in the basis of the electron orbital states and present it as a sum of
two terms
\begin{equation}
H_\mscr{HF}=H^0_\mscr{HF}+V_\mscr{HF},
\end{equation}
where the terms are defined as
\begin{eqnarray}
H^0_\mscr{HF}&=&A\sum_l | l\rangle \langle l| \hat{\bm{S}}\cdot
\sum_k |\langle \bm{R}_k|l \rangle |^2 \hat{\bm{I}}_k 
\label{eq:H0hyperfine}\\
V_\mscr{HF}&=&\sum_{l\neq l'}
| l\rangle \langle l'| \hat{\bm{S}}\cdot
\sum_k \langle l | \bm{R}_k \rangle \langle \bm{R}_k|l' \rangle
\hat{\bm{I}}_k.
\label{eq:Vhyperfine}
\end{eqnarray}
By definition $H^0_\mscr{HF}$ does not couple different orbital levels.  
By combining Eqs.\ (\ref{eq:electronHamiltonian})
and (\ref{eq:H0hyperfine}) one obtains the following Hamiltonian
\begin{equation}
H_0=\sum_l 
\biglb (
\varepsilon_l +g\mu_B \hat{\bm{\mathcal{B}}}_l \cdot \hat{\bm{S}}
\bigrb )|l\rangle \langle l|.
\label{eq:H0final}
\end{equation}
We will regard the mixing term $V_\mscr{HF}$ as a perturbation to $H_0$. 
The justification for this is that
the typical fluctuations of the electron energy due to the
hyperfine interaction are much smaller than the orbital energy
separation, $\hbar \Omega$.

\begin{figure}[ht]
\includegraphics[angle=0,width=5cm]{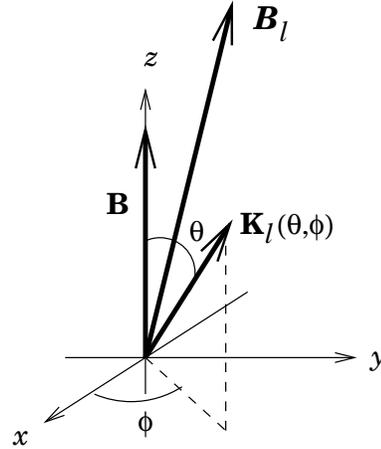}
\caption{The internal field coordinate system is set by the external
magnetic field $\bm{B}$, i.e.\ $\bm{e}_z \parallel \bm{B}$.
The combination of the external and the internal field $\bm{K}_l$
results in an effective field $\mathcal{B}_l$.}
\label{fig:figure1}
\end{figure}

We now concentrate on $H_0$ and formulate semiclassical description of
it. For this we consider the operator of the orbitally dependent effective
magnetic field;
\begin{equation}
\hat{\bm{\mathcal{B}}}_l=\bm{B}+\frac{1}{g\mu_B}\hat{\bm{K}}_l
\end{equation}
where 
\begin{equation}
\hat{\bm{K}}_l=\frac{E_n}{I C_n}\sum_k |\langle \bm{R}_k|l \rangle |^2 \hat{\bm{I}}_k .
\label{eq:Koperator}
\end{equation}
Our goal is to replace the operator $\hat{\bm{K}}_l$ with a classical
field. 
To prove the replacement is reasonable we calculate the average of the
square for a given unpolarized nuclear
state $ |\bm{\mu} \rangle$
\begin{equation}
\bm{K}^2_l=\langle \bm{\mu}
|\hat{\bm{K}}_l^2|\bm{\mu}\rangle=\frac{E_n^2}{N_\mscr{QD}}  
\rho_l^{[2]} (I+1)/I
\end{equation}
where $\rho_l^{[n]}=\int d\bm{r} V_\mscr{QD}^{n-1} 
|\langle \bm{r}|l\rangle|^{2n}$ is a dimensionless constant 
depending on the wave function of the orbital $l$. 
We cannot simply replace $\hat{\bm{K}}_l$ by its eigenvalues:
as in the case of the usual spin algebra, different components of
$\hat{\bm{K}}_l$ 
do not commute.  In addition its square does not commute with
individual components, $[\hat{\bm{K}}_l^2,K_l^\alpha] \neq 0$. 
To estimate fluctuations of $\hat{\bm{K}}_l$ we calculate the
uncertainty relations between its components
\begin{eqnarray}
\Delta K_l^\alpha \Delta K_l^\beta \ge
\frac{E_n^2}{N_\mscr{QD}^{3/2}} 
=\bm{K}_l^2  \frac{1}{N_\mscr{QD}^{1/2}}.
\end{eqnarray}
Since $N_\mscr{QD}^{-1/2} \ll 1$ we have proved that the quantum 
fluctuations in $\hat{\bm{K}}_l$ are much smaller than its typical
amplitude.   
The semiclassical picture introduced above is only valid for high
temperatures, $kT \gg E_n
N_\mscr{QD}^{-1}$, where there are many states available to the
nuclear system and the typical length is proportional to
$N_\mscr{QD}^{1/2}$.  For temperatures below $E_n   
N_\mscr{QD}^{-1}$ the nuclear system will predominantly be in the
ground state and the classical picture breaks down.  
This is similar to the quantum mechanical description
of a particle moving in a potential.  At zero temperature it will
be localized in some potential minimum and quantum mechanics will 
dominate. At sufficiently high temperatures the particle
occupies higher energy states and its motion is
well described by classical mechanics. 
Having established this we can replace the average over the density
matrix of the nuclei by the average over a classical field $\bm{K}_l$,
note that it has no `hat', whose values are Gaussian distributed
\begin{equation}
P(\bm{K}_l)=\left (\frac{3}{2 \pi \langle \bm{K}_l^2\rangle}
\right )^{3/2}  
\exp \Biglb (-\frac{3 (\bm{K}_l-\bm{K}_l^{(0)})^2}
                   {2 \langle \bm{K}_l^2 \rangle} \Bigrb ).
\label{eq:Kweight}
\end{equation}
Here $\bm{K}_l^{(0)}$ is the average, or Overhauser, field assuming
small degree of polarization.
The effective magnetic field acting on the electron is 
$\bm{\mathcal{B}_l}=\mathcal{B}_l\bm{n}$ where $\bm{n}$ is the unit vector
along the total field for a given $\bm{K}_l$, see Fig.\
\ref{fig:figure1}.  For this configuration 
the spin eigenfunctions are $| \bm{n}_\pm\rangle$ corresponding to
eigenvalues 
\begin{equation}
\bm{n}\cdot \hat{\bm{S}} | \bm{n}_\pm\rangle= \pm \frac{1}{2}|
\bm{n}_\pm\rangle 
\label{eq:doublet}
\end{equation}
The effective Zeeman Hamiltonian is then
\begin{equation}
H_\mscr{Z}=g \mu_B \bm{\mathcal{B}}_l\cdot \hat{\bm{S}}
\end{equation}
in this given internal field configuration.  
The spectrum of $H_0$ thus consists of many doublets distinguished by
the value of $\bm{\mathcal{B}}_l$.
The magnitude of the
effective field $\mathcal{B}_l$ 
determines the Zeeman splitting of the each doublet
\begin{eqnarray}
\Delta_l&=&g \mu_B \mathcal{B}_l \nonumber \\
&=&(E_B^2+K_l^2+2E_B K_l \cos \theta)^{1/2} 
\label{eq:zeemanSplitting}
\end{eqnarray}
where $E_B=g\mu_B B$ the external field Zeeman energy.

We conclude this section with two remarks concerning time and
energy scales. 
First, since the dynamics of $\bm{K}_l$ is essentially the dynamics of the
nuclear system it changes on a time scale of the nuclear relaxation,
which can be quite long.\cite{berg90:2563}
For electron dynamics on shorter timescales $\bm{K}_l$ plays the
role of a `frozen disorder'.  At longer time scales self-averaging
over all configurations of $\bm{K}_l$
takes place.
Second, the RMS value of $\bm{K}_l$ is approximately $5$\,$\mu$eV
which corresponds to a 100\,Gauss field (for a value $N_\mscr{QD}
\approx 10^5$).
\section{Transition Rate}
\label{sectionIV}
We concentrate on the transitions between the doublet components in Eq.\
(\ref{eq:doublet}).
Assuming that the higher energy doublet state is initially occupied,
we will calculate the transition rate to the lower one.
The transition must be accompanied by 
energy dissipation equal to $\Delta_0$. 
This energy cannot be absorbed be the nuclear
system.\cite{kim94:16777,erlingsson01:195306} 
So we need an external mechanism of energy dissipation.
The most effective one is known to be phonons. However the phonons
alone cannot change the electron spin so we need a mechanism which
mixes spin and orbital degrees of freedom, that is $V_\mscr{HF}$ from
Eq.\ (\ref{eq:Vhyperfine}).   
Thus the transition amplitude is
proportional to both $V_\mscr{HF}$ and the electron  phonon coupling
$H_\mscr{ph}$.

Here we assume that the electron is in the lowest orbital state
$| 0 \rangle$ since the phonon mechanism will bring the electron to
this state from any  
higher orbital on timescales much smaller than those related to
transtions between the doublet components.
Thus we consider an initial state of the entire system
$|\mbox{i}\rangle=|0,\bm{n}_-;\bm{\mu};\bm{N}\rangle$ which is a
product state of the 
electron, nuclear $|\bm{\mu}\rangle$ and phonon $|\bm{N}\rangle$ 
systems and the final state
$|\mbox{f}\rangle=|0,\bm{n}_+;\bm{\mu}';\bm{N}'\rangle$.
The transition amplitude between them, in second order perturbation
theory, reads
\begin{eqnarray}
\mathcal{T}&=&\sum_{l \neq 0}
\left ( 
\frac{
\langle 0,\bm{n}_+;\bm{\mu}'|V_\mscr{HF}| l,\bm{n}_-;\bm{\mu}\rangle
\langle l;\bm{N}'|H_\mscr{ph}|0;\bm{N} \rangle 
}{
(\varepsilon_0-\varepsilon_l)+E_B}
\right. \nonumber \\
& &+
\left .
\frac{
\langle 0;\bm{N}'|H_\mscr{ph}|l;\bm{N} \rangle
\langle l,\bm{n}_+;\bm{\mu}'|V_\mscr{HF}| 0,\bm{n}_-;\bm{\mu}\rangle
}{
(\varepsilon_0-\varepsilon_l)-E_B}
\right).
\label{eq:transitionAmplitude}
\end{eqnarray}
The summation is over virtual states involving higher orbitals and the 
denominators in Eq.\ (\ref{eq:transitionAmplitude}) contains
the energy differences between different orbital states.  The internal
field depends on the orbital state, resulting in a rather complicated
expression.
Albeit the energy related to the internal field is much smaller than
the orbital separation so we can safely replace 
the Zeeman splitting with $E_B$. The reason for that is that only at
high external fields where $\Delta_0 \approx E_B$ will the effects of
the Zeeman splitting be appreciable in the denominator.
The internal field also appears in the phonon rate
since it determines the electron energy difference  between the
initial and final state.
Since $H_\mscr{ph}$ does not connect different nuclear states and 
conversely $V_\mscr{HF}$ does not mix different phonon states the sums
over intermediate phonon and nuclear states reduce to a single term.  
From this transition amplitude the transition rate is obtained via
Fermis golden rule
\begin{equation}
\tilde{\Gamma}_\mscr{sf}=\frac{2\pi}{\hbar}\sum_{\bm{N}'}
\sum_{\bm{\mu}'}|\mathcal{T}|^2 \delta(E_\mscr{i}-E_\mscr{f}),
\label{eq:fermisGoldenRule}
\end{equation}
where $E_\mscr{i}-E_\mscr{f}$ is the energy difference
between the initial and final state of the combined systems.
Substituting Eq.\ (\ref{eq:transitionAmplitude}) into Eq.\
(\ref{eq:fermisGoldenRule}) we get the following relation for the
spin-flip rate
\begin{widetext}
\begin{eqnarray}
\tilde{\Gamma}_\mscr{sf}&=&
\sum_{l,l' \neq 0} 
\left(
\frac{
\langle l,\bm{n}_-;\bm{\mu}|V_\mscr{HF}|0,\bm{n}_+\rangle
\langle 0,\bm{n}_+|V_\mscr{HF}|l',\bm{n}_-;\bm{\mu} \rangle
}{
((\varepsilon_0-\varepsilon_l)+E_B)
((\varepsilon_0-\varepsilon_{l'})+E_B) 
}
\frac{2\pi}{\hbar}\sum_{\bm{N}'}
\langle 0;\bm{N}|H_\mscr{ph}| l;\bm{N}'\rangle
\langle l';\bm{N}'|H_\mscr{ph}| 0;\bm{N}\rangle
\delta(E_\mscr{i}-E_\mscr{f})
\right . \nonumber \\
&&\hspace{0.6cm}+
\frac{
\langle l,\bm{n}_-;\bm{\mu}|V_\mscr{HF}|0,\bm{n}_+\rangle
\langle l',\bm{n}_+|V_\mscr{HF}|0,\bm{n}_-;\bm{\mu} \rangle
}{
((\varepsilon_0-\varepsilon_l)+E_B)
((\varepsilon_0-\varepsilon_{l'})-E_B) 
}
\frac{2\pi}{\hbar}\sum_{\bm{N}'}
\langle 0;\bm{N}|H_\mscr{ph}| l;\bm{N}'\rangle
\langle 0;\bm{N}'|H_\mscr{ph}| l';\bm{N}\rangle
\delta(E_\mscr{i}-E_\mscr{f})
\nonumber \\
&&\hspace{0.6cm}+
\frac{
\langle 0,\bm{n}_-;\bm{\mu}|V_\mscr{HF}|l,\bm{n}_+\rangle
\langle 0,\bm{n}_+|V_\mscr{HF}|l',\bm{n}_-;\bm{\mu} \rangle
}{
((\varepsilon_0-\varepsilon_l)-E_B)
((\varepsilon_0-\varepsilon_{l'})+E_B) 
}
\frac{2\pi}{\hbar}\sum_{\bm{N}'}
\langle l;\bm{N}|H_\mscr{ph}| 0;\bm{N}'\rangle
\langle l';\bm{N}'|H_\mscr{ph}| 0;\bm{N}\rangle
\delta(E_\mscr{i}-E_\mscr{f})
\nonumber \\
&&\hspace{0.6cm}+
\left.
\frac{
\langle 0,\bm{n}_-;\bm{\mu}|V_\mscr{HF}|l,\bm{n}_+\rangle
\langle l',\bm{n}_+|V_\mscr{HF}|0,\bm{n}_-;\bm{\mu} \rangle
}{
((\varepsilon_0-\varepsilon_l)-E_B)
((\varepsilon_0-\varepsilon_{l'})-E_B) 
}
\frac{2\pi}{\hbar}\sum_{\bm{N}'}
\langle l;\bm{N}|H_\mscr{ph}| 0;\bm{N}'\rangle
\langle 0;\bm{N}'|H_\mscr{ph}|l';\bm{N}\rangle
\delta(E_\mscr{i}-E_\mscr{f})
\right ).
\label{eq:transitionRate}
\end{eqnarray}
\end{widetext}
The spin-flip rate depends on the initial state of the nuclear system
$|\bm{\mu}\rangle$.  This poses the problem of how to deal with the
nuclear state $|\bm{\mu}\rangle$, since we already demoted all
the spin operators to a collective classical variable.  A
conceptually simple solution lies in the fact that when Eq.\
(\ref{eq:transitionRate}) and (\ref{eq:Vhyperfine}) 
are considered together one sees that the rate is a sum over all pairs
of nuclei in the system.  Focusing on a given pair of nuclei $k$
and $k'$, all the other nuclei are unchanged when the electron spin  is
`scattered' on by this pair.  By simply redefining the classical
field such that it is composed of all nuclei except this given pair we
can circumvent the problem.
This procedure will not 
change our previous result regarding the properties of $\bm{K}_l$ and
also defining $|\bm{\mu}\rangle =| m_k\rangle |m_{k'} \rangle$ 
makes it straightforward to work with the nuclear states in Eq.\
(\ref{eq:transitionRate}).

Although the transition rate can be very slow, the typical duration of
a transition event is set by energy uncertainty, $\hbar/\Delta_0$.
This is much shorter than the typical time for nuclear spin relaxation
so that the  nuclear system is frozen in
a given configuration of $\bm{K}_0$.
In this case taking an average over $\bm{K}_0$
is not well motivated since a given configuration of $\bm{K}_0$ is
fixed.  For now we will postpone the average over the classical
field.   
After performing a thermal average over $|\bm{\mu}\rangle$ in the
transition rate we obtain the following
\begin{widetext}
\begin{eqnarray}
\Gamma_\mscr{sf}&=& G_\mscr{corr} \left (
\sum_{l\neq 0} \left \{ \frac{2 a_{ll}
\gamma_{ll}}{\delta \varepsilon_l^2}
\left (1+3 \frac{E_B^2}{\delta
\varepsilon_l^2 } \right )
+\frac{2 \Re\{\tilde{a}_{ll}
\tilde{\gamma}_{ll}\}}{\delta \varepsilon_l^2}
\left (1+\frac{E_B^2}{\delta
\varepsilon_l^2}  \right) 
\right \} \right .\nonumber \\
& &\hspace{-2.0cm}\left. +
\sum_{l<l'\neq 0} \left \{
\frac{4 \Re\{a_{ll'}\gamma_{ll'} \}}
     {\delta \varepsilon_l \delta \varepsilon_{l'}} 
\Biglb (1+ \Biglb ( \frac{(\delta \varepsilon_l +\delta \varepsilon_{l'})^2}
                         {\delta \varepsilon_l \delta \varepsilon_{l'}}
                   -1 \Bigrb ) 
                  \frac{E_B^2}{\delta \varepsilon_l \delta
\varepsilon_{l'}}\Bigrb )  
+
\frac{4 \Re\{\tilde{a}_{ll'}\tilde{\gamma}_{ll'} \}}
     {\delta \varepsilon_l \delta \varepsilon_{l'}} 
\Biglb (1+ \Biglb ( \frac{(\delta \varepsilon_l -\delta \varepsilon_{l'})^2}
                         {\delta \varepsilon_l \delta \varepsilon_{l'}}
                   +1 \Bigrb ) 
                  \frac{E_B^2}{\delta \varepsilon_l \delta
\varepsilon_{l'}}\Bigrb)  
\right \}
\right )
\label{eq:transitionRateExpanded}
\end{eqnarray}
\end{widetext}
where $\delta \varepsilon_l=\varepsilon_0-\varepsilon_l$.  
Here we have introduced the one-site nuclear spin correlation function 
$G_\mscr{corr}$, which is evaluated in appendix \ref{app:Gcorr}.
The denominator has been expanded to second order in the Zeeman
splitting.
The parameters $a_{ll'},\tilde{a}_{ll'}$ are related to the $V_\mscr{HF}$
matrix elements  
\begin{eqnarray}
a_{ll'}\!&=&\! A^2C_n \int d^3\bm{R}_k \Psi_l^*(\bm{R}_k)
|\Psi_0(\bm{R}_k)|^2
\Psi_{l'}(\bm{R}_k) 
\label{eq:a}\\
\tilde{a}_{ll'}\!&=&\! A^2C_n \int d^3\bm{R}_k \Psi_l^*(\bm{R}_k)
\Psi_{l'}^*(\bm{R}_k) 
\Psi_0(\bm{R}_k)^2
\label{eq:atilde}
\end{eqnarray}
and $\gamma_{ll'},\tilde{\gamma}_{ll'}$ are generalized phonon
transition rates 
\begin{eqnarray}
\gamma_{ll'}\!\!&=&\!\! \frac{2\pi}{\hbar} \!\! \sum_{\bm{q} \nu}
\frac{\alpha^2_\nu(\bm{q}) 
[e^{-i\bm{q}\cdot \bm{r}}]_{0,l} [e^{i\bm{q}\cdot \bm{r}}]_{l',0}}
{e^{\beta \hbar \omega_{\bm{q},\nu}}-1}\delta(\hbar
\omega_{\bm{q}\nu}\!\!-\!\!\Delta_0)
\label{eq:gamma}  \\
\tilde{\gamma}_{ll'}\!\!&=&\!\! \frac{2\pi}{\hbar}\!\!\sum_{\bm{q} \nu}
\frac{\alpha^2_\nu(\bm{q}) 
[e^{-i\bm{q}\cdot \bm{r}}]_{0,l} [e^{i\bm{q}\cdot \bm{r}}]_{0,l'}}
{e^{\beta \hbar \omega_{\bm{q},\nu}}-1}\delta(\hbar
\omega_{\bm{q}\nu}\!\!-\!\! \Delta_0).
\label{eq:gammatilde}
\end{eqnarray}
Here we have only included the emission process since we assume the
the spin is initially in the higher energy state.  

Until now we have considered a general quantum dot and the rate in
Eq.\ (\ref{eq:transitionRateExpanded}) is valid for any quantum dot.  To
proceed further we will specify the confining potential to be
parabolic in the lateral direction and in the transverse one a
triangular well potential is chosen.   
The wave function is $\langle \bm{r}|l \rangle \equiv \chi_0(z)
\psi_{n,M}(r,\theta)$ where $n,M$ denote the orbital and angular
momentum quantum numbers respectively of the Darwin-Fock solution and
$\chi_0(z)$ is the wave function in the transverse direction. 
The square of the lateral confining length is $\ell^2=\hbar^2/ m^*
\hbar \Omega$, where $m^*$ is the electron effective mass and
$\Omega=(\Omega_0^2+(\omega_c/2)^2)^{1/2}$ is the effective confining
frequency.  The confining energy is 
$\hbar \Omega_0$ and $\omega_c=eB/m^*$ is the cyclotron frequency.  
What remains is to calculate the $a$'s and $\gamma$'s in Eqs.\
(\ref{eq:a})-(\ref{eq:gammatilde}).  The results of these calculations
are presented appendix \ref{app:integral}.

\begin{figure}[ht]
\includegraphics[angle=-90,width=8.5cm]{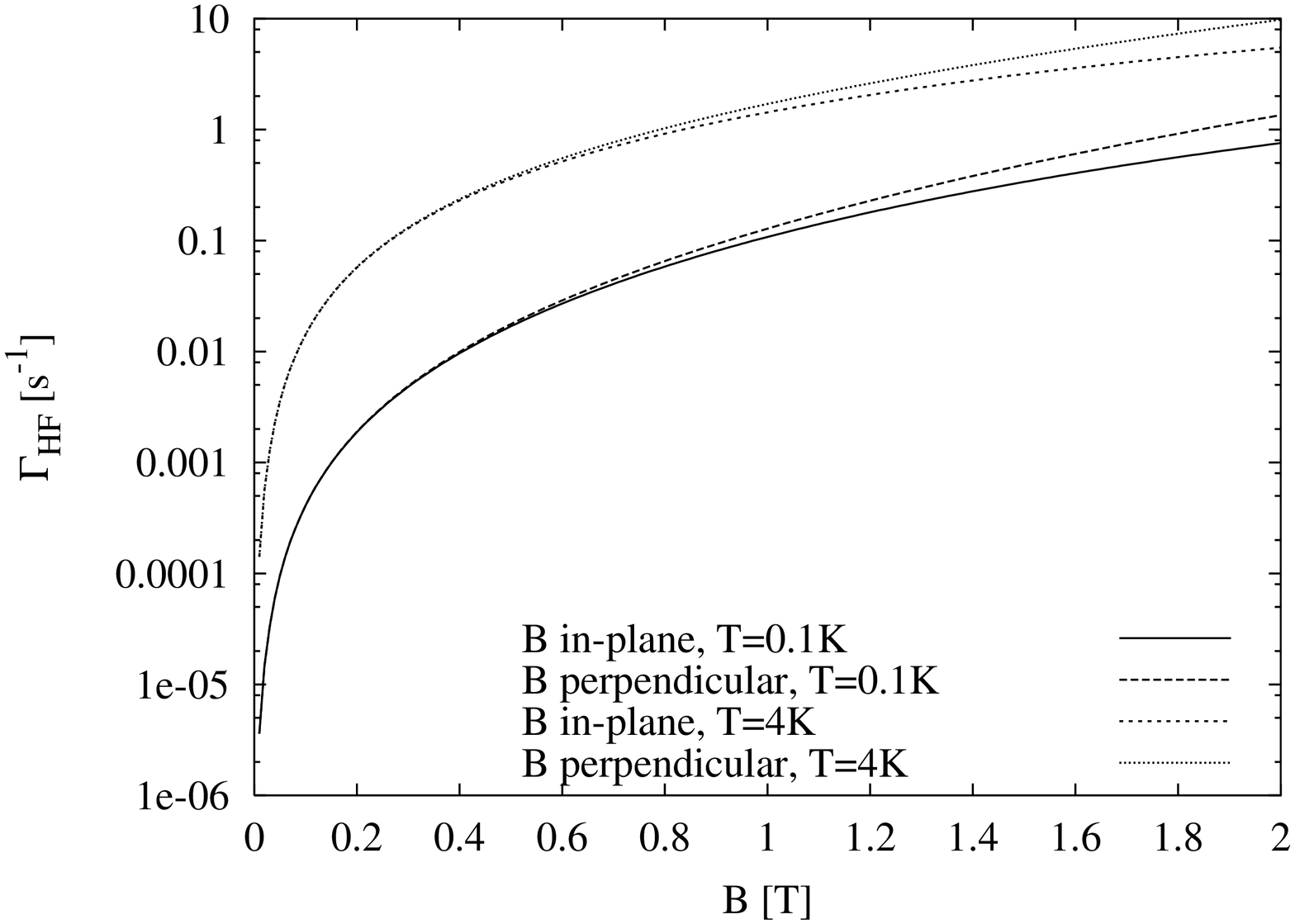}
\caption{The hyperfine-mediated spin-flip rate for a quantum dot with
$z_0=10$\,nm and $\hbar \Omega_0=5$\,meV, plotted a function of
external magnetic field for two different temperatures $T=0.1$\,K and
4\,K.}
\label{fig:figure2}
\end{figure}

\begin{figure}[ht]
\includegraphics[angle=-90,width=8.5cm]{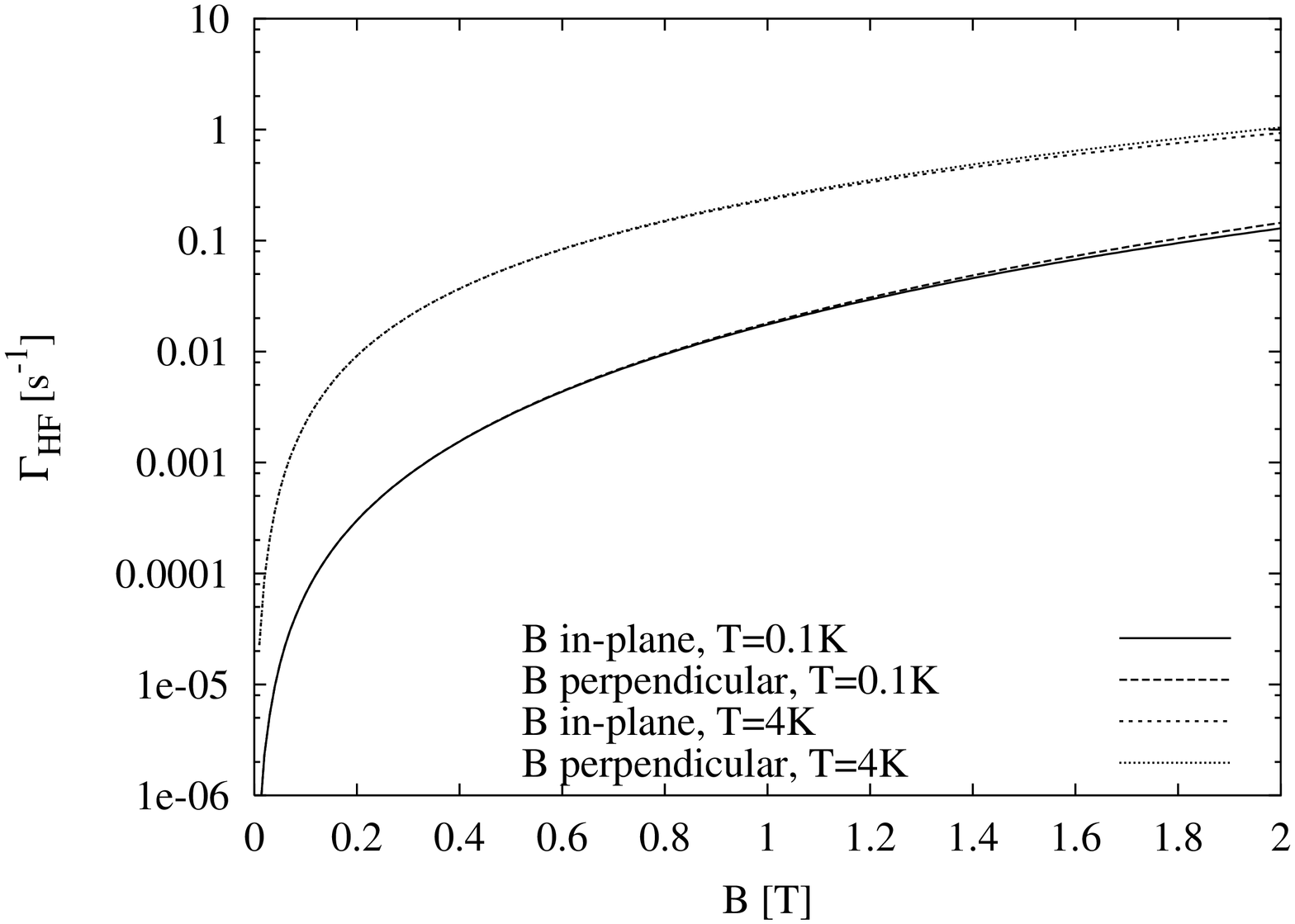}
\caption{The hyperfine-mediated spin-flip rate for a quantum dot with
$z_0=10$\,nm and $\hbar \Omega_0=5$\,meV, plotted a function of
external magnetic field for two different temperatures $T=0.1$\,K and
4\,K.}
\label{fig:figure3}
\end{figure}

In principle it is possible to obtain the rate
for all parameter values but to make the discussion more transparent
we will consider regimes of the applies magnetic field $(i)$ $E_B \approx
E_n N_\mscr{QD}^{-1/2}$ and $(ii)$ $E_B \gg E_n N_\mscr{QD}^{-1/2} $.     
In regime $(i)$ both $\Delta_0 \ll
\hbar c_\nu \ell^{-1}$ and $\Delta_0 \ll kT$ (for experimentally
relevant temperatures) and only the lowest order terms in 
$\Delta_0/\hbar c_\nu \ell^{-1}$ need to be considered.  
In GaAs $\hbar c_l \ell^{-1}=3.3\ell^{-1}$\,nm$\times$meV and $\hbar c_t
\ell^{-1}=2.0\ell^{-1}$\,nm$\times$meV
for the longitudinal and transverse branches respectively. 
In the other regime we assume that the applied field dominates and that the
internal field may be ignored, but no additional assumption are made
in this case.  This will in general require some numerical work.  The
resulting hyperfine-mediated spin-flip rates are
\begin{widetext}
\begin{eqnarray}
\Gamma_\mscr{HF}&=&0.34\frac{G_\mscr{corr}}{I^2}
\frac{E_n^2}{N\mscr{QD}(\hbar \Omega)^2}
\frac{(eh_{14}\ell)^2 kT }
{8\pi \rho c^5 \hbar^4}
\left (
E_B^2+K_0^2+2E_B K_0 \cos \theta
\right ) 
\mbox{\hspace{0.3cm}for\,\,} E_B \approx E_n N_\mscr{QD}^{-1/2}
\label{eq:finalRateLow} \\
\Gamma_\mscr{HF}&=&\frac{G_\mscr{corr}}{I^2}
\frac{E_n^2(n(E_B)+1)}{N\mscr{QD}(\hbar \Omega)^2}
\frac{(eh_{14}\ell)^2 E_B^3}
{8\pi \rho c^5 \hbar^4}
\left (
C_0(E_B)+\left (\frac{E_B}{\hbar \Omega} \right)^2
C_2(E_B) 
\right) \mbox{\hspace{0.3cm}for\,\,} 
E_B \gg E_n N_\mscr{QD}^{-1/2} 
\label{eq:finalRate}
\end{eqnarray}
\end{widetext}
Note that the rates have different dependence on the
emitted energy $\Delta_0$.  In regime $(i)$ it  is proportional to
$\Delta_0^2$ due to the Bose statistics, and in regime
$(ii)$ it is proportional to $\Delta_0^3$ for $\Delta_0 \ll
\hbar c_\nu \ell^{-1}$.  
In Eq.\ (\ref{eq:finalRate}) we introduce the functions $C_0$
and $C_2$ which contain the details of the higher orbitals and the anisotropy
integrals.  For low fields $\Delta_0 \ll \hbar c_\nu \ell^{-1}$ these
functions are constant.  
The value for the spin flip rate in Eq.\ (\ref{eq:finalRateLow}), for
some typical value of $K_0 \approx 10^{-2}$\,T, is very low $\Gamma_\mscr{HF}
\approx 10^{-6}$\,s.  This results in a lifetime of days, which will
be extremely difficult to measure.  
We have plotted the general spin-flip rate in Eq.\
(\ref{eq:finalRate}) for different confining energies 
and temperatures in Figs.\
\ref{fig:figure2} and \ref{fig:figure3}, for both in-plane and
perpendicular applied magnetic field.  In the case of the higher
confining energy there is very small difference between the in-plane
and perpendicular direction of the external magnetic field.  For the
lower confining energy there is a substantial difference between the
two directions of magnetic field.  In this case the approximation
$\Delta_0 \ll \hbar c_\nu \ell^{-1}$ is no longer good and the the
$\Delta_0^3$ dependence of the rate is changed by the $C$-functions. 
The values of the rates are quite small, depending on the applied
field, being in $\sim 1$\,s${^{-1}}$ for $T=4$\,K at $B \approx
0.5$\,T for a confining frequency of $\hbar \Omega_0=2$\,meV. 
\section{Discussion}
Generally speaking spin-flip rates require external dissipation.  This
is why at small Zeeman splittings they will contain small factor
reflecting the vanishing phonon density of states.
For a spin-orbit rate\cite{khaetskii01:27}, the Kramer's degeneracy
results in this small factor being proportional to  $E_B^5$.
Presence of nuclear spins violate the Kramer's theorem.
So that the hyperfine rate discussed in the present paper 
is proportional to $E_B^3$ and will dominate at sufficiently low fields.

Comparing the hyperfine rate in Eq.\ (\ref{eq:finalRate}) to
spin-orbit related rates\cite{khaetskii01:27} $\Gamma_\mscr{SO}$ we
obtain that the hyperfine rate dominates for fields below $B \approx 0.3$\,T
(assuming $z_0=10$\,nm and $\hbar \Omega_0=4$\,meV). 

The role of the internal field produced by the nuclei is that the
spin-flip rate does not vanish even for in the absence of external
magnetic field. The show that the minimum rate is rather small,
corresponding to a relaxation time of the order of days.
We believe that the internal field  will play an important role when the full
dynamics of the electron spin in the presence of the nuclear system is
considered. 

Our model should also be applicable to other polar semiconductors
which have non-zero nuclear spin, e.g.\ InAs where the $g$-factor is
much larger.
\label{sectionV}

\section{Aknowledgement}
One author (SIE) would like to thank Daniela Pfannkuche, Alexander
Chudnovskii and Daniel Loss for fruitful discussions. This work is a 
part of the research program of the ``Stichting vor Fundementeel Onderzoek der
Materie (FOM)'' 
\appendix
\section{Average over nuclear spin pair}
\label{app:Gcorr}
When the thermal average over a pair of free nuclei the following 
nuclear correlation function appear in Eq.\ (\ref{eq:transitionRateExpanded})
\begin{eqnarray}
G_\mscr{corr}&=&(S_{+-}^\alpha)^*S_{+-}^\beta 
\langle \delta I^\alpha_{k} \delta I^\beta_{k'} \rangle_T
\nonumber \\
&=&(S_{+-}^\alpha)^*S_{+-}^\beta 
\langle \delta \hat{I}^\alpha \delta \hat{I}^\beta\rangle_T \delta_{k,k'},
\end{eqnarray}
where $\delta \hat{I}^\alpha=\hat{I}^\alpha-\langle I^\alpha
\rangle_T$ and the electron spin matrix elements being
$S_{+-}^\alpha=\langle \bm{n}_+ |S^\alpha| \bm{n}_- \rangle$. 
The kronekker delta reflects that there are no correlations between
two different nuclei and we have dropped the $k$ subscript in $\langle
\delta \hat{I}^\alpha \delta \hat{I}^\beta\rangle_T$ since the nuclei is assumed
to be identical.  By defining the symmetric correlator 
\begin{equation}
g^{\alpha \beta}=\frac{1}{2} \langle \delta \hat{I}^\alpha \delta \hat{I}^\beta + \delta
\hat{I}^\beta \delta \hat{I}^\alpha \rangle_T
\end{equation}
we get the following
\begin{equation}
G_\mscr{corr}=(S_{+-}^\alpha)^*S_{+-}^\beta (g^{\alpha \beta}+i/2
\epsilon^{\alpha \beta \gamma} \langle \hat{I}^\gamma\rangle_T).
\end{equation}
In an isotropic system, $\langle \hat{\bm{I}} \rangle_T=0$, the
value of the correlation function is
$G_\mscr{corr}=\frac{1}{2}\frac{1}{3} I(I+1)$. 
\section{Matrix elements for a parabolic quantum dot}
\label{app:integral}
Using the Darwin-Fock solutions and the Fang-Howard variational
solution for the triangular quantum well we obtain the following 
equation are Eqs. (\ref{eq:a})-(\ref{eq:gammatilde})  
\begin{widetext}
\begin{eqnarray}
a_{ll'}\gamma_{ll'}&=&
\delta_{M,M'}
\frac{A^2Cn}{V_\mscr{QD} \zeta}
\frac{(eh_{14}\ell)^2\Delta^3}{8\pi \rho c^5 \hbar^4} 
(n(\Delta)+1) 
\frac{\Gamma(n+n'+|M|+1)2^{-3(n+n'+|M|)}}
{n! n'! (n+|M|)!(n'+|M|)!}
\nonumber \\
& &\times
\left (
\sum_\nu \frac{c^5}{c_\nu^5} 
\left (\frac{\Delta\ell}{\hbar c_\nu} \right)^{2(n+n'+|M|-1)}
\int_0^\pi d(\cos \theta)  A_\nu(\theta)
\frac{
(\sin \theta)^{2(n+n'+|M|)} 
\exp \Biglb (-\frac{1}{2} \biglb ( \frac{\Delta \sin
\theta}{\hbar c_\nu \ell^{-1}} \bigrb )^2 \Bigrb) 
}
{
\left (
1+ \left ( \frac{\Delta\cos \theta}{3 \hbar c_\nu z_0^{-1}} \right)^2
\right )^3
}
\right )
\label{eq:agamma}
\end{eqnarray}
\end{widetext}
where $\zeta^{-1}=z_0\int dz |\chi(z)|^4$ and
$c^{-5}=c_l^{-5}+c_t^{-5}$ is the effective sound velocity of the
phonons.  The anisotropy functions are
\begin{eqnarray}
A_t(\theta)&=&\frac{\sin^2 \theta (8 \cos^4 \theta+\sin^4 \theta)}{4} \\
A_l(\theta)&=&\frac{9 \cos^2 \theta \sin^4 \theta}{2}.
\end{eqnarray}
The equation for $\tilde{a}_{ll'}\tilde{\gamma}_{ll'}$ is identical
except for a different Kronekker delta function $\delta_{M,-M'}$.
The integral needs in general to be evaluated numerically but when
$\Delta \ll \hbar c_\nu \ell^{-1}$ the exponential term and the
denominator become unity and the resulting integral is simple to
calculate. 
\end{document}